\documentclass[aps,pra,nobibnotes,11pt]{revtex4-1}
\usepackage{amsmath}
\usepackage{graphicx}

\usepackage{array}
\newcommand{\PreserveBackslash}[1]{\let\temp=\\#1\let\\=\temp}
\newcolumntype{C}[1]{>{\PreserveBackslash\centering}p{#1}}
\newcolumntype{R}[1]{>{\PreserveBackslash\raggedleft}p{#1}}
\newcolumntype{L}[1]{>{\PreserveBackslash\raggedright}p{#1}}

\begin{document}

\title[O. M. G. Ward and E. McCann]{The heat equation for nanoconstrictions in 2D materials with Joule self-heating}

\author{Oliver M. G. Ward}
\address{Physics Department, Lancaster University, Lancaster, LA1 4YB, UK}

\author{Edward McCann}
\address{Physics Department, Lancaster University, Lancaster, LA1 4YB, UK}\email{ed.mccann@lancaster.ac.uk}

\begin{abstract}
We consider the heat equation for monolayer two-dimensional materials in the presence of heat flow into a substrate and Joule heating due to electrical current. We compare devices including a nanowire of constant width and a bow tie (or wedge) constriction of varying width, and we derive approximate one-dimensional heat equations for them; a bow tie constriction is described by the modified Bessel equation of zero order. We compare steady state analytic solutions of the approximate equations with numerical results obtained by a finite element method solution of the two-dimensional equation. Using these solutions, we describe the role of thermal conductivity, thermal boundary resistance with the substrate and device geometry.
The temperature in a device at fixed potential difference will remain finite as the width shrinks, but will diverge for fixed current, logarithmically with width for the bow tie as compared to an inverse square dependence in a nanowire.
\end{abstract}

\maketitle

\section{Introduction}

There is huge interest in the thermal properties of two-dimensional (2D) materials, motivated by applications to thermal management~\cite{pop10,fu20,lewis21,sachat21},
interconnects in integrated circuits~\cite{ferrari15,debroy20,son21},
thermoelectric devices~\cite{zong20,wang20,pallecchi20,zhao21} and nanoscale fabrication~\cite{jeong14,island14,abbassi17,gu18}. While nanoscale constrictions are of great importance for their electrical transport characteristics, particularly in the quantum regime~\cite{celis16,saraswat21}, they also promote enhanced Joule self-heating and thermoelectric coefficients~\cite{liao11,zolotavin17,harzheim18,evangeli21}.
For such structures, scanning thermal microscopy~\cite{sachat21} is an ideal tool to map surface temperatures with spatial resolution of a few nanometres, as applied to carbon nanotubes~\cite{shi09}, nanowires~\cite{puyoo11,menges16,sachet17,gachter20} and 2D materials~\cite{pumarol12,menges13,yoon14,tortello16,choi17,yalon17,harzheim18,tortello19,sachat19,yasaei19,evangeli19,harzheim20,evangeli21}.

In this paper, we model heat transport and the resulting spatial temperature profile in the presence of Joule self-heating due to electrical current in nanoconstrictions in monolayer 2D materials. We consider electrical and thermal transport in the classical, diffusive regime, applicable to experiments at room temperature or above, and with spatial dimensions typically of the order of~$100\,$nm or above~\cite{liao11,dorgan13,zolotavin17,harzheim18,evangeli21}. In this regime, modelling typically involves analytic solution of the one-dimensional heat equation~\cite{durkan99,pop05,kuroda05,pop07,shi09,dorgan13,hunley13,chandran15,pu19} or numerical solutions using the finite element method~\cite{fangohr11,islam13,ramos15,harzheim18}. 

\begin{figure}[t]
\includegraphics[scale=0.5]{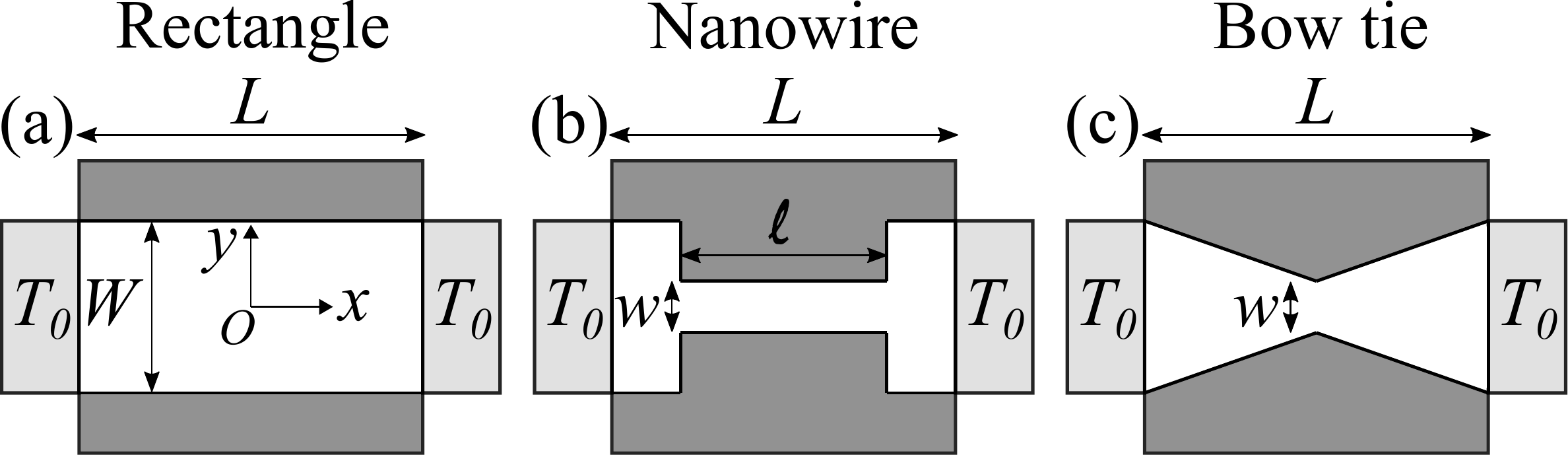}
\caption{Schematic plan view of three different two-lead devices with (a) a uniform rectangle of length $L$ and width $W$, (b) a nanowire constriction of length $\ell$ and constant width $w$, (c) a bow tie constriction of minimum width $w$. The central white region indicates the device as described by the heat equation~(\ref{he2}). Light grey indicates two leads at temperature $T_0$ with boundary condition~(\ref{bcx}), dark grey indicates insulating regions with boundary condition~(\ref{bcy}). Cartesian coordinate axes $(x,y)$ are shown in (a) with the origin $O$ at the centre of the sample.
}\label{fig1}
\end{figure}

We consider three geometries as typical examples, Fig.~\ref{fig1}, which are a rectangle, a nanowire~\cite{hadeed07,liao11,puyoo11,xiang14,menges16,sachet17,zolotavin17,sawtelle19,gachter20} of constant width $w$, and a bow tie (or wedge) constriction~\cite{ramachandran05,lu10,harzheim18,sawtelle19,evangeli21} of varying width down to a minimum $w$. We compare the temperature profiles for devices with the same macroscopic dimensions and characteristic parameters, for either a fixed applied potential difference or a fixed current.
The nanowire and bow tie are excellent representative examples because they exhibit markedly different behaviour: the electrical resistances of the nanowire and bow tie have different dependences on the width $w$, and this means that Joule heating is independent of $w$ for the nanowire (for small $w$ and fixed potential difference), but diverges as $w \rightarrow 0$ for the bow tie.
We derive effective one-dimensional heat equations for the three systems with analytic solutions, and we compare them with numerical solutions using the finite element method~\cite{fem}.
This approach highlights the central role played by the thermal healing length $L_{\mathrm{H}}$ which encapsulates sample specific details about the thermal conductivity and the rate of heat loss to the environment including a substrate~\cite{chiang02,pop07}.

We consider the steady state heat equation in two dimensions~\cite{hahn13,hahncomment1}
\begin{eqnarray}
\nabla \cdot \left( \kappa \nabla T({\bf r}) \right) + p({\bf r}) = 0 , \label{he1}
\end{eqnarray}
where $p({\bf r})$ is the areal rate of internal energy generation (power per unit area)~\cite{dimensionscomment} and $\kappa$ is the homogeneous thermal conductivity.
We include two opposing sources of energy generation $p({\bf r}) = p_{\mathrm{J}}({\bf r}) + p_{\mathrm{B}}({\bf r})$ where $p_{\mathrm{J}}({\bf r}) = \mathbf{j} \cdot \mathbf{E}$ describes Joule heating with electrical current density $\mathbf{j}$ and electric field $\mathbf{E}$,
and $p_{\mathrm{B}}({\bf r})$ describes heat loss from the 2D monolayer sample in the vertical direction through sample-substrate or sample-air interfaces (depending on the particular setup).
This may be parameterised with a thermal boundary resistance $R_{\mathrm{B}}$ such that $p_{\mathrm{B}}({\bf r}) = - [T({\bf r}) - T_0]/R_{\mathrm{B}}$ for ambient temperature of the environment $T_0$~\cite{durkan99,chiang02,pop05,kuroda05,dorgan10,balandin11}.
We assume that the sample and the substrate are homogeneous, so that the thermal boundary resistance $R_{\mathrm{B}}$ is independent of position over the whole area of the sample (the central white regions in Fig.~\ref{fig1}).

Thus, the heat equation~(\ref{he1}) may be written as
\begin{eqnarray}
\nabla \cdot \left( \kappa \nabla T({\bf r}) \right) +  \mathbf{j} \cdot \mathbf{E} - \frac{[T({\bf r}) - T_0]}{R_{\mathrm{B}}} = 0 , \label{he1a}
\end{eqnarray}
Comparison of the first term with the other two shows that
\begin{eqnarray}
T({\bf r}) \approx
\begin{cases}
T_0 + T_{\mathrm{J}}({\bf r}) , &\mbox{for } \kappa \rightarrow 0 , \\
T_0 , & \mbox{for } \kappa \rightarrow \infty ,
\end{cases}  \label{lhlimit}
\end{eqnarray}
where $\kappa$ should be compared with $L^2/R_{\mathrm{B}}$ for sample dimension $L$. Here $T_{\mathrm{J}}({\bf r}) = p_{\mathrm{J}}({\bf r})  R_{\mathrm{B}} = \mathbf{j} \cdot \mathbf{E} R_{\mathrm{B}}$ is the `Joule heating temperature' which is a product of the Joule heating power per unit area $\mathbf{j} \cdot \mathbf{E}$
and the thermal boundary resistance $R_{\mathrm{B}}$. Joule heating is the local source of temperature increase that is generally dependent on position ${\bf r}$ as determined by conservation of current and the sample shape (we study the examples in Fig.~\ref{fig1} in detail later).
The heat equation~(\ref{he1a}) describes how the sample temperature $T({\bf r})$ is related to the inhomogeneous source term $T_{\mathrm{J}}({\bf r})$.
For a poor thermal conductor, Eq.~(\ref{lhlimit}) shows that the temperature increase at position ${\bf r}$ is actually equal to the Joule heating temperature $T_{\mathrm{J}}({\bf r})$, whereas an excellent thermal conductor remains close to the ambient temperature $T_0$.

We write the heat equation~(\ref{he1a}) solely in terms of functions and parameters with dimensions of temperature or length,
\begin{eqnarray}
\nabla^2 T({\bf r}) + \frac{T_0 + T_{\mathrm{J}}({\bf r}) - T({\bf r})}{L_{\mathrm{H}}^2} = 0 , \label{he2}
\end{eqnarray}
where $L_{\mathrm{H}} = \sqrt{\kappa R_{\mathrm{B}}}$ is the thermal healing length~\cite{chiang02,kuroda05,pop07}.
Eq.~(\ref{he2}) illustrates that solutions $T({\bf r})$ will be a function of ${\bf r}/L_{\mathrm{H}}$, i.e. the thermal healing length $L_{\mathrm{H}}$ is the length scale of typical variations of the temperature (the limits in Eq.~(\ref{lhlimit}) could be written with $L_{\mathrm{H}} = \sqrt{\kappa R_{\mathrm{B}}}$ instead of $\kappa$ if $R_{\mathrm{B}}$ is kept finite).
Intermediate $L_{\mathrm{H}}$ means that different regions of a device cannot be considered as separate components with individual, isolated thermal resistances, and this will often be the case, e.g. $L_{\mathrm{H}}$ is estimated to be of the order of $100\,$ nm in graphene~\cite{pop12,choi17}, WTe$_2$~\cite{mleczko16} and MoS$_2$~\cite{yalon17} (its value depends on the materials and sample quality).

We consider a two-lead set up with connections to electrical leads in the longitudinal $x$ direction at $x = \pm L/2$ for a sample of length $L$, Fig.~\ref{fig1}. We neglect thermal boundary resistance at the contacts~\cite{pop10,sachat21,barcohen15,denisov20} and consider the leads to be excellent heat sinks fixed at the ambient temperature $T_0$ as described by Dirichlet boundary conditions,
\begin{eqnarray}
\left. T \right|_{\substack{\text{longitudinal} \\ \text{boundary}}} = T_0 . \label{bcx}
\end{eqnarray}
In the transverse $y$ direction, we consider connection of the sample to a thermal insulator with a Neumann boundary condition,
\begin{eqnarray}
\left. \nabla T \cdot \mathbf{\hat{n}} \right|_{\substack{\text{transverse} \\ \text{boundary}}} = 0 , \label{bcy}
\end{eqnarray}
where $\mathbf{\hat{n}}$ is the normal to the boundary.
In the following, we solve the heat diffusion equation~(\ref{he2}) for systems with different geometries, Fig.~\ref{fig1}, forms of Joule heating $T_{\mathrm{J}}({\bf r})$, and values of $L_{\mathrm{H}}$.

\section{Rectangular sample}

We consider an homogeneous rectangular sample of length $L$ and width $W$ with Cartesian coordinates $(x,y)$, $-L/2 \leq x \leq L/2$, $-W/2 \leq y \leq W/2$, Fig.~\ref{fig1}(a). Electrical current $I$ flows along the $x$ direction in response to a potential difference $V$.
The power per unit area due to Joule heating can be written as $p_{\mathrm{J}} = \mathbf{j} \cdot \mathbf{E} = P/(LW)$ where the power is $P = IV$, and temperature $T_{\mathrm{J}} = P R_B/(LW)$ is independent of position.
Owing to Joule heating and the boundary conditions~(\ref{bcx}), there is a spatially-dependent temperature profile $T({\bf r})$, where $T({\bf r}) > T_0$ with the external environment at ambient temperature $T_0$.
In the linear response regime, $V= IR$ for electrical resistance $R$. For the rectangular device, we denote this as $R_{\mathrm{rect}} = L / (\sigma W)$ for dc conductivity $\sigma$.
The Joule heating temperature is
\begin{eqnarray}
T_{\mathrm{J},\mathrm{rect}} = \frac{I^2 R_{\mathrm{B}}}{\sigma W^2}
= \frac{V^2 \sigma R_{\mathrm{B}}}{L^2} .
\label{trect}
\end{eqnarray}
For fixed electrical current $I$, $T_{\mathrm{J},\mathrm{rect}}$ increases as $W$ decreases (because the current density increases), but, for a fixed potential difference $V$, $T_{\mathrm{J},\mathrm{rect}}$ is proportional to the power per unit area, and it is independent of $W$,
i.e. $I \propto W$, but current density $I/W$ is independent of $W$.

\begin{figure}[t]
\includegraphics[scale=0.39]{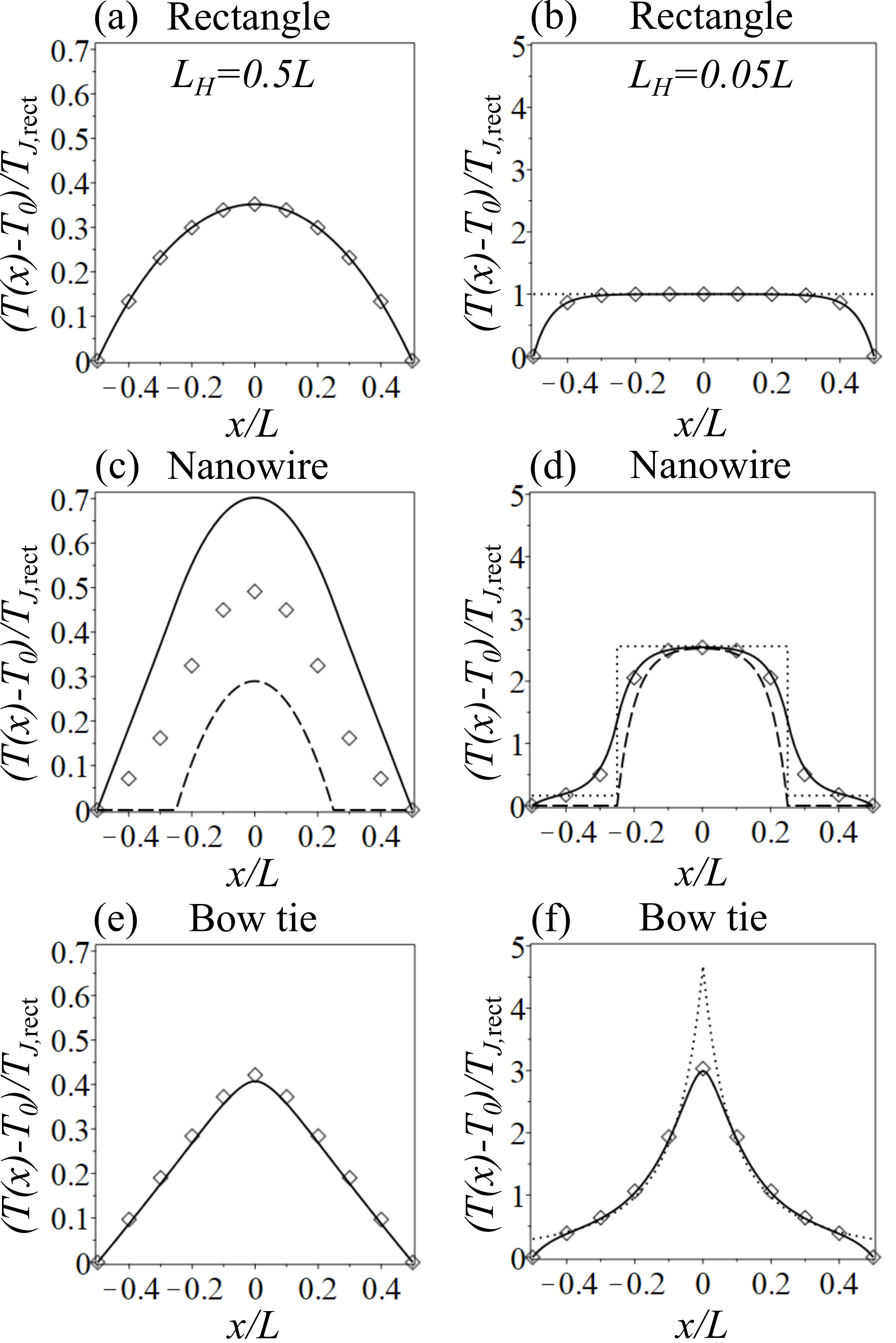}
\caption{The temperature profile $T(x) - T_0$ in the longitudinal direction normalized by the Joule temperature for a rectangle $T_{\mathrm{J},\mathrm{rect}} = V^2 \sigma R_{\mathrm{B}} / L^2$.
The first column (a), (c), (e) is for $L_{\mathrm{H}} = 0.5L$, the second column (b), (d), (f) is for $L_{\mathrm{H}} = 0.05L$~\cite{rbcomment}.
The first row (a), (b) is for a rectangular device, the second row (c), (d) for a nanowire, and the third row (e), (f) for a bow tie. For all devices, $W = 0.5L$. The nanowire length is $\ell = 0.5L$, and the nanowire and bow tie constriction have minimum width $w = 0.125L$. Solid lines show the temperature found from one-dimensional equations~(\ref{rtemp},\ref{twire},\ref{twedge}) for the rectangle, nanowire and bow tie, respectively. Dashed lines in (c),(d) show the additional solution~(\ref{model1}) for the nanowire. Diamond symbols show numerical data from the finite element method solution~\cite{fem} of the two-dimensional equation~(\ref{he2}), and dotted lines in the second column (b), (d), (f) show the Joule heating profile~(\ref{lhlimit}) that is valid in the limit $L_{\mathrm{H}} \rightarrow 0$.
Note the different vertical scales in the two columns.
}\label{fig2}
\end{figure}

We apply separation of variables~\cite{hahncomment3} to the two-dimensional heat equation~(\ref{he2}).
With boundary conditions~(\ref{bcy}) in the transverse direction at $y = \pm W/2$, the $y$ dependence of the temperature is constant, reducing the heat equation~(\ref{he2}) to an 
effective one-dimensional equation~\cite{chiang02,pop05,dorgan13},
\begin{eqnarray}
\frac{d^2 T (x)}{dx^2} + \frac{T_0 + T_{\mathrm{J}} - T(x)}{L_{\mathrm{H}}^2} = 0 . \label{rhe}
\end{eqnarray}
We assume that $\kappa$, $\sigma$ and $R_{\mathrm{B}}$ are independent of temperature within the range of temperatures $T(x)$. Then, the general solution of Eq.~(\ref{rhe}) is given by exponentials,
\begin{eqnarray}
T(x) = T_0 + T_{\mathrm{J}} + A e^{x / L_{\mathrm{H}}} + B e^{- x / L_{\mathrm{H}}} , \label{rgs}
\end{eqnarray}
where $A$ and $B$ are arbitrary constants.
With the boundary conditions~(\ref{bcx}), $T(x=-L/2) = T(x=L/2) = T_0$,
the temperature~\cite{chiang02,kuroda05,pop07,pop10,liao10,chandran15} is
\begin{eqnarray}
T (x) = T_0 + T_{\mathrm{J}} \!\left[ 1 - \frac{\cosh (x/L_{\mathrm{H}})}{\cosh (L/(2L_{\mathrm{H}}))} \right] . \label{rtemp}
\end{eqnarray}
This temperature profile is plotted in Fig.~\ref{fig2}(a),(b) for different values of $L_{\mathrm{H}}$ (solid lines). The maximum temperature is at the centre of the sample, $T (0) = T_0 + T_{\mathrm{J}} - T_{\mathrm{J}}/\cosh (L/(2L_{\mathrm{H}}))$, and
\begin{eqnarray}
T (0) \approx
\begin{cases}
T_0 , &\mbox{for } \kappa \gg L^2/R_{\mathrm{B}}  , \\
T_0 + T_{\mathrm{J}} ,
 & \mbox{for } \kappa \ll L^2/R_{\mathrm{B}} ,
\end{cases} \label{rlimits}
\end{eqnarray}
where $T_{\mathrm{J}} \equiv T_{\mathrm{J},\mathrm{rect}} = V^2 \sigma R_{\mathrm{B}} / L^2$ for the rectangle. These limits agree with the initial expectations~(\ref{lhlimit}): the constant profile of Joule heating $T = T_0 + T_{\mathrm{J},\mathrm{rect}}$ is shown as a dotted line in Fig.~\ref{fig2}(b).
When taking the limits in Eq.~(\ref{rlimits}) we keep $R_{\mathrm{B}}$ fixed. Otherwise, the dependence of the Joule heating temperature $T_{\mathrm{J}} = V^2 \sigma R_{\mathrm{B}} / L^2$ on $R_{\mathrm{B}}$ should also be taken into account (see table~\ref{tablevi}).

We compare our analytical results with numerical solution of the two-dimensional heat equation~(\ref{he2}) using the finite element method~\cite{fem}. Numerical data is shown as symbols in Fig.~\ref{fig2}. For the rectangular geometry, the analytic solution~(\ref{rtemp}) is exact, as shown by the agreement of analytical and numerical data in Fig.~\ref{fig2}(a),(b).

\section{Nanowire}

For a uniform rectangle with a fixed value of potential difference $V$, the characteristic Joule heating temperature $T_{\mathrm{J}} \equiv T_{\mathrm{J},\mathrm{rect}} = V^2 \sigma R_{\mathrm{B}} / L^2$ doesn't depend on the sample width $W$. To better understand the dependence on sample width, we 
consider a nanowire of length $\ell$ and constant width $w$ at the centre of the device of total length $L$ and width $W$, Fig.~\ref{fig1}(b). We assume that the material parameters $\kappa$, $\sigma$ and $R_{\mathrm{B}}$ are homogeneous thoughout the sample.
To begin, we will determine the Joule heating temperature in each region.

Adding classical resistors in series gives the total electrical resistance as
\begin{eqnarray}
R_{\mathrm{wire}} = \frac{Lw + \ell (W-w)}{\sigma w W} , \label{rwire}
\end{eqnarray}
and $R_{\mathrm{wire}} > R_{\mathrm{rect}}$ for devices with the same bulk dimensions $L$, $W$ and material parameters.
For an applied potential difference $V$, and recalling that the current $I = V / R_{\mathrm{wire}}$ is constant along the device, we find that the Joule heating temperatures, $T_{\mathrm{J},w}$ and $T_{\mathrm{J},W}$, in the regions of width $w$ and $W$, respectively, are given by
\begin{eqnarray}
T_{\mathrm{J},w} &=& \frac{I^2 R_{\mathrm{B}}}{\sigma w^2}
= \frac{V^2 \sigma W^2 R_{\mathrm{B}}}{\left[ Lw + \ell (W-w) \right]^2} , \label{Tw} \\
T_{\mathrm{J},W} &=& \frac{I^2 R_{\mathrm{B}}}{\sigma W^2}
= \frac{V^2 \sigma w^2 R_{\mathrm{B}}}{\left[ Lw + \ell (W-w) \right]^2} . \label{TW}
\end{eqnarray}
Thus the power dissipated per unit area in the nanowire increases as $T_{\mathrm{J},w} / T_{\mathrm{J},W} = ( W / w )^2$ when the width drops from $W$ to $w$.
However, for $\ell \neq 0$, the total resistance $R_{\mathrm{wire}} \rightarrow \ell / (\sigma w)$ diverges as $w \rightarrow 0$, and $T_{\mathrm{J},w} \rightarrow V^2 \sigma R_{\mathrm{B}} /\ell^2$ is the same as that of a rectangle of length $\ell$, width $w$, and it is independent of $w$ in this limit (for fixed $V$).
The profile of the Joule heating temperature in the longitudinal direction, $T_0 + T_{\mathrm{J}}(x)$, is shown as the dotted line in Fig.~\ref{fig2}(d) for $W/w = 4$, and this is equal to the temperature $T(x)$ in the limit $L_{\mathrm{H}} \rightarrow 0$ (except for a small region close to the boundary).

Separation of variables~\cite{hahncomment3} is not easily applicable to the nanowire geometry, Fig.~\ref{fig1}(b), because the heat equation and boundary conditions are inhomogeneous~\cite{pu19}. We find two approximate analytic solutions, one underestimates the temperature, the other overestimates. The first solution is found by assuming that the parts of the device of width $W$ have negligible electrical resistance for $w \ll W$ (or $L - \ell \ll L$) and simply form part of the external leads at temperature $T_0$. Then, the central nanowire of width $w$, $|x| \leq \ell/2$, can be considered as a rectangle of length $\ell$ and width $w$ as in Eq.~(\ref{rtemp}),
\begin{eqnarray}
T (x) \approx T_0 + T_{\mathrm{J},w} \!\left[ 1 - \frac{\cosh (x/L_{\mathrm{H}})}{\cosh (\ell/(2L_{\mathrm{H}}))} \right] . \label{model1}
\end{eqnarray}
This is shown as the dashed line in Fig.~\ref{fig2}(c),(d) and it generally underestimates the correct $T(x)$ (symbols) because it neglects self-heating in the parts of width $W$.

The second approximate solution is found by assuming that the temperature in each rectangular region (of width $w$ or $W$) is described by the general solution~(\ref{rgs}) of a rectangle (i.e. it is independent of $y$).
We then match $T(x)$ and $dT/dx$ at the boundaries $x = \pm \ell / 2$ between the separate rectangular regions to give
\begin{eqnarray}
T(x) \approx \begin{cases}
T_0 + T_{\mathrm{J},w} - \tilde{T}_{\mathrm{J}} \cosh \left( \frac{x}{L_{\mathrm{H}}} \right) , &\mbox{for } - \frac{\ell}{2} \leq x \leq \frac{\ell}{2} , \\
T_0 + T_{\mathrm{J},W} \left[ 1 - \cosh \left( \frac{L - 2|x|}{2L_{\mathrm{H}}} \right) \right]
+ {\cal C}^{-1} \! \left[ T_{\mathrm{J},W} {\cal S} + \tilde{T}_{\mathrm{J}} \sinh \left( \frac{\ell}{2L_{\mathrm{H}}} \right) \right] \! \sinh \left( \frac{L - 2|x|}{2L_{\mathrm{H}}} \right) \! ,
 & \mbox{for } \frac{\ell}{2} \leq |x| \leq \frac{L}{2} ,
\end{cases} \nonumber \\ \label{twire}
\end{eqnarray}
where
\begin{eqnarray*}
\tilde{T}_{\mathrm{J}} &=& \frac{T_{\mathrm{J},w} {\cal C} + T_{\mathrm{J},W} (1 - {\cal C})}{{\cal C} \cosh (\ell / (2L_{\mathrm{H}}) ) + {\cal S}\sinh (\ell / (2L_{\mathrm{H}}) )} , \\
{\cal C} &=& \cosh \left( \frac{L - \ell}{2L_{\mathrm{H}}} \right) \, ; \qquad
{\cal S} = \sinh \left( \frac{L - \ell}{2L_{\mathrm{H}}} \right) .
\end{eqnarray*}
Eq.~(\ref{twire}) has the same spatial dependence as Eq.~(\ref{model1}) in the central region, $|x| \leq \ell / 2$, but the form of the parameter $\tilde{T}_{\mathrm{J}}$ is different.
The temperature profile~(\ref{twire}) is plotted in Fig.~\ref{fig2}(c),(d) for different values of $L_{\mathrm{H}}$ (solid lines), and compared with numerical data (symbols) from the finite element method solution~\cite{fem} of the two-dimensional equation~(\ref{he2}).
The solution~(\ref{twire}) is quite inaccurate when $L_{\mathrm{H}}$ is of the order of the system dimensions, Fig.~\ref{fig2}(c), although it is very accurate for small $L_{\mathrm{H}}$, Fig.~\ref{fig2}(d).
Solution~(\ref{twire}) generally overestimates $T(x)$ because,
by assuming the temperature has no $y$ dependence,
it neglects heat spreading into a larger area in the parts of width $W$. As shown in Fig.~\ref{fig2}(c), the numerical data points (symbols) lie between the two analytical estimates.

\begin{figure}[t]
\includegraphics[scale=0.38]{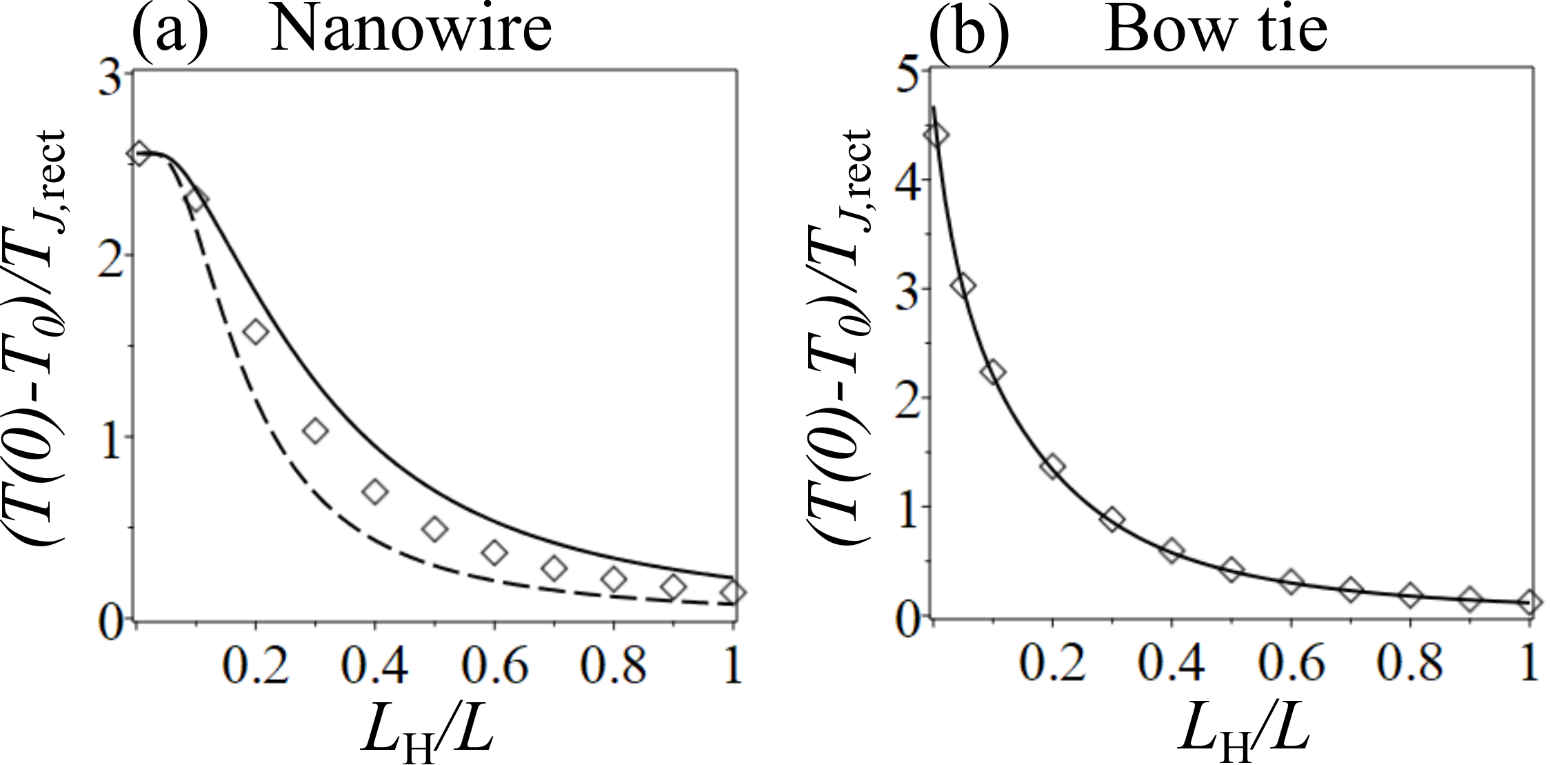}
\caption{The maximum temperature $T(0) - T_0$ as a function of the healing length
$L_{\mathrm{H}}$~\cite{rbcomment}
for minimum width $w = 0.125L$ where diamond symbols show numerical data from the finite element method solution~\cite{fem} of the two-dimensional equation~(\ref{he2}). The plots are normalized by the Joule temperature for a rectangle $T_{\mathrm{J},\mathrm{rect}} = V^2 \sigma R_{\mathrm{B}} / L^2$.
(a) is for a nanowire of length $\ell = 0.5L$, and the dashed and solid lines show the temperature found from the one-dimensional equations~(\ref{model1},\ref{twire}), respectively.
(b) is for a bow tie, and the solid line shows the temperature found from one-dimensional equation~(\ref{twedge}). For both plots, $W = 0.5L$. Note the different vertical scales in (a) and (b).
}\label{fig3}
\end{figure}

The maximum temperature is at the centre of the sample, $T(0)$, and is determined by the spatial dimensions of the device. As shown in Fig.~\ref{fig2}(a),(c), the nanowire geometry doesn't result in a much greater temperature than the rectangle for large $L_{\mathrm{H}}$: enhancement occurs for smaller $L_{\mathrm{H}}$ as illustrated in Fig.~\ref{fig2}(b),(d), in line with the original prediction~(\ref{lhlimit}).
Both of the analytic models, Eqs.~(\ref{model1},\ref{twire}), give
\begin{eqnarray}
T (0) \approx
\begin{cases}
T_0 , &\mbox{for } \kappa \gg L^2/R_{\mathrm{B}}  , \\
T_0 + T_{\mathrm{J},w} ,
 & \mbox{for } \kappa \ll L^2/R_{\mathrm{B}} ,
\end{cases} \label{lhlimit2}
\end{eqnarray}
where $T_{\mathrm{J},w}$ is the Joule heating temperature of the nanowire section with narrow width $w$, Eq.~(\ref{Tw}).
Dependence of $T(0)$ as a function of
$L_{\mathrm{H}}$~\cite{rbcomment}
is shown in Figure~\ref{fig3}(a) where the symbols show numerical data~\cite{fem}, the dashed line is Eq.~(\ref{model1}) and the solid line is Eq.~(\ref{twire}).

\begin{figure}[t]
\includegraphics[scale=0.38]{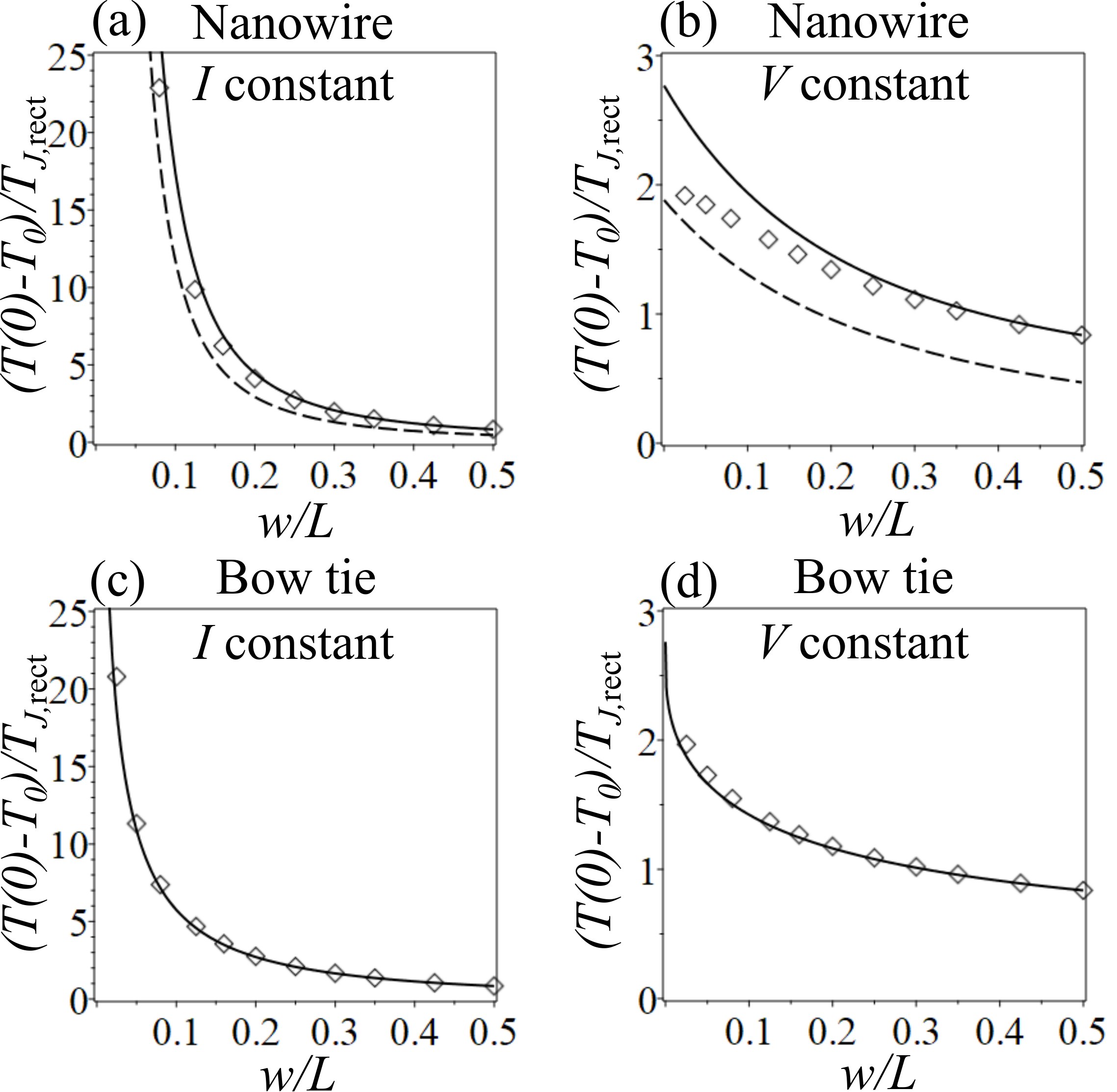}
\caption{The maximum temperature $T(0) - T_0$ as function of the minimum width $w$ for $L_{\mathrm{H}} = 0.2L$ where diamond symbols show numerical data from the finite element method solution~\cite{fem} of the two-dimensional equation~(\ref{he2}). The plots are normalized by the Joule temperature for a rectangle $T_{\mathrm{J},\mathrm{rect}}$, Eq.~(\ref{trect}). The first row (a), (b) is for a nanowire of length $\ell = 0.5L$, and the dashed and solid lines show the temperature found from the one-dimensional equations~(\ref{model1},\ref{twire}), respectively.
The second row (c), (d) is for a bow tie constriction, and the solid lines show the temperature found from the one-dimensional equation~(\ref{twedge}).
The first column (a), (c) is for constant electrical current $I$, and the second column (b), (d) is for constant potential difference $V$.
For all plots, $W = 0.5L$.
}\label{fig4}
\end{figure}

The dependence of $T(0)$ as a function of width $w$ is shown in Fig.~\ref{fig4}(a) for constant current and in Fig.~\ref{fig4}(b) for constant potential difference $V$.
Symbols show numerical data~\cite{fem}, the dashed line is Eq.~(\ref{model1}) and the solid line is Eq.~(\ref{twire}).
In both plots, Eq.~(\ref{twire}) is a good approximation for $w \alt W$, and, for $w = W$, the device is simply a rectangle, Eq.~(\ref{rtemp}), with $T (0) = T_0 + T_{\mathrm{J}} - T_{\mathrm{J}}/\cosh (L/(2L_{\mathrm{H}}))$ and $T_{\mathrm{J}} = V^2 \sigma R_{\mathrm{B}} / L^2$.
For constant current, Fig.~\ref{fig4}(a), the temperature diverges as $w \rightarrow 0$ because the current density diverges.
For constant potential difference, Fig.~\ref{fig4}(b), the temperature is finite for $w \rightarrow 0$. The electrical resistance of the part of width $W$ is negligible
(for $w \rightarrow 0$),
and Eq.~(\ref{model1}) gives $T (0) \approx T_0 + T_{\mathrm{J},w} - T_{\mathrm{J},w}/\cosh (\ell/(2L_{\mathrm{H}}))$ with $T_{\mathrm{J},w} \approx V^2 \sigma R_{\mathrm{B}} / \ell^2$.

Results for the maximum temperature $T(0)$ in various limits are summarised in table~\ref{tablevi}. Dependence on $\kappa$ follows the expected behaviour~(\ref{lhlimit},\ref{lhlimit2}) (a poor thermal conductor leads to an increased temperature). Although the thermal boundary resistance $R_{\mathrm{B}}$ also appears in the thermal healing length $L_{\mathrm{H}} = \sqrt{\kappa R_{\mathrm{B}}}$, it appears in the Joule heating temperature $T_{\mathrm{J}}$, too. Hence, the dependence on $R_{\mathrm{B}}$ is distinct from that of $\kappa$: when $R_{\mathrm{B}} \rightarrow 0$, heat can immediately dissipate into the environment and $T(0) \rightarrow T_0$, whereas $R_{\mathrm{B}} \rightarrow \infty$ increases the temperature $T(0)$ at the centre of the device.

\begin{table}[h]
\begin{center}
\caption{\label{tablevi}Parameter dependence of the maximum excess temperature $T(0) - T_0$ where $T(0)$ is the maximum temperature (at $x = 0$) and $T_0$ is the ambient temperature. Limits are determined using Eq.~(\ref{model1}) for the nanowire and Eq.~(\ref{tzbt}) for the bow tie. The top rows are for fixed voltage $V$, the bottom rows are for fixed current $I$, and we use $f (\ell) =  \left[ 1 - \frac{1}{\cosh (\ell/(2\sqrt{\kappa R_{\mathrm{B}}}))}\right]$. Results for the rectangle can be obtained from those for the nanowire with the substitutions $w \rightarrow W$ and $\ell \rightarrow L$.}
\begin{tabular}{ L{1.7cm} | C{2.3cm} | C{1.8cm} | C{2.3cm} | C{1.8cm} | C{2.3cm} | C{2.3cm} }
\hline \hline
fixed$\,\,V$ & $\kappa \rightarrow 0$ & $R_{\mathrm{B}} \rightarrow 0$ & $w \rightarrow 0$
& $\kappa \rightarrow \infty$ & $R_{\mathrm{B}} \rightarrow \infty$ & $w = W, \ell = L$ \\ 
\hline
nanowire & $\frac{V^2 \sigma W^2 R_{\mathrm{B}}}{[Lw + \ell (W-w)]^2}$ & $0$ & $\frac{V^2 \sigma R_{\mathrm{B}} f  (\ell)}{\ell^2}$ & $0$ & $\frac{V^2 \sigma W^2 \ell^2}{8\kappa [Lw + \ell (W-w)]^2}$ & $\frac{V^2 \sigma R_{\mathrm{B}} f (L)}{L^2}$  \\  
bow tie & $\frac{V^2 \sigma (W-w)^2 R_{\mathrm{B}}}{w^2L^2 \ln^2(W/w)}$ & $0$ & $\frac{V^2 \sigma}{8\kappa}$ & $0$ & $\frac{V^2 \sigma}{8\kappa}$ & $\frac{V^2 \sigma R_{\mathrm{B}} f (L)}{L^2}$ \\
\hline \hline
fixed$\,\,I$& $\kappa \rightarrow 0$ & $R_{\mathrm{B}} \rightarrow 0$ & $w \rightarrow 0$
& $\kappa \rightarrow \infty$ & $R_{\mathrm{B}} \rightarrow \infty$ & $w = W, \ell = L$ \\ 
\hline
nanowire & $\frac{I^2 R_{\mathrm{B}}}{\sigma w^2}$ & $0$ & $\frac{I^2 R_{\mathrm{B}} f (\ell)}{\sigma w^2}$ & $0$ & $\frac{I^2 \ell^2}{8\sigma \kappa w^2}$ & $\frac{I^2 R_{\mathrm{B}}f (L)}{\sigma W^2}$  \\  
bow tie & $\frac{I^2 R_{\mathrm{B}}}{\sigma w^2}$ & $0$ & $\frac{I^2 L^2 \ln^2 (W/w)}{8 \sigma \kappa W^2}$ & $0$ & $\frac{I^2 L^2 \ln^2 (W/w)}{8 \sigma \kappa (W-w)^2}$ & $\frac{I^2 R_{\mathrm{B}}f (L)}{\sigma W^2}$ \\
\hline \hline
\end{tabular}
\end{center}
\end{table}

\section{Bow tie constriction}

\begin{figure}[t]
\includegraphics[scale=0.60]{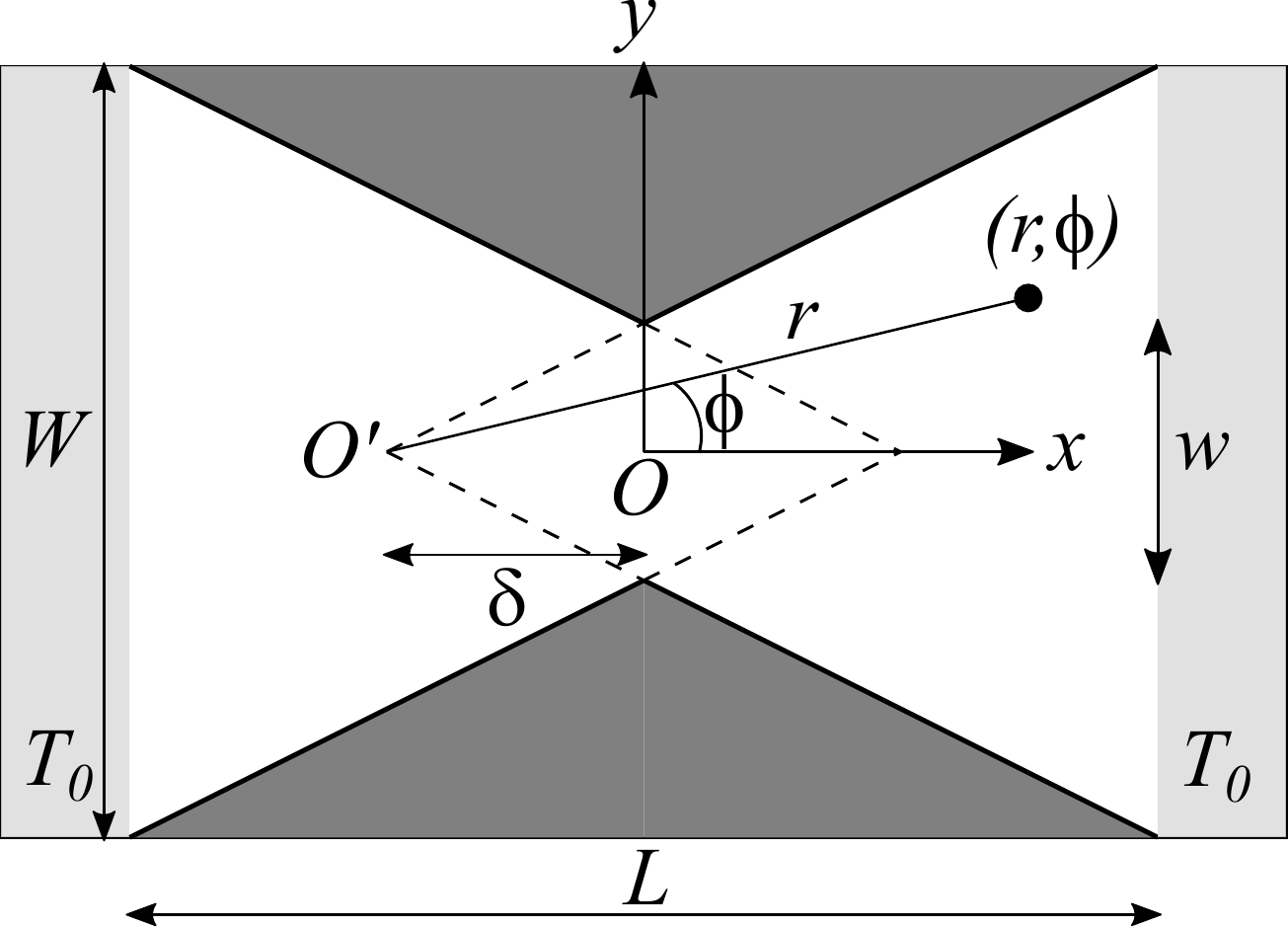}
\caption{Schematic plan view of a bow tie constriction of minimum width $w$, maximum width $W$ and total length $L$. The central white region indicates the device as described by the heat equation~(\ref{he2}). Light grey indicates two leads at temperature $T_0$ with boundary condition~(\ref{bcx}), dark grey indicates insulating regions with boundary condition~(\ref{bcy}). Cartesian coordinate axes $(x,y)$ are shown with the origin $O$ at the centre of the sample. The sloping sides of the bowtie meet at the virtual vertex $O^{\prime}$ which serves as the origin for polar coordinates $(r, \phi)$, and is offset by distance $\delta$ from $O$.
}\label{newfig1}
\end{figure}

We consider a bow tie constriction, Fig.~\ref{fig1}(c) and Fig.~\ref{newfig1}, with length $L$ and width ${\cal W} (x) = w + 2 |x| (W - w)/L$ that varies from  ${\cal W} (x=0) = w$ at the centre to ${\cal W} (x=\pm L/2) = W$ at the edge, $W > w$. For a strip at $x$ with width ${\cal W} (x)$ and infinitesimally short length $\Delta x$, the resistance is $\Delta R = \Delta x / (\sigma {\cal W} (x))$. Summing classical resistances (i.e. integrating with respect to $x$) gives the total electrical resistance of the bow tie system as
\begin{eqnarray}
R_{\mathrm{bowtie}} = \frac{L}{\sigma (W-w)} \ln \! \left( \frac{W}{w} \right) , \label{rbowtie}
\end{eqnarray}
and $R_{\mathrm{bowtie}} > R_{\mathrm{rect}}$ for devices with the same bulk dimensions $L$, $W$ and material parameters.
The current $I = V / R_{\mathrm{bowtie}}$ is constant along the device, although the current density increases as the constriction narrows. Thus we find that the Joule heating temperature is inhomogeneous as
\begin{eqnarray}
T_{\mathrm{J},\mathrm{bowtie}} (x)
= \frac{I^2 R_{\mathrm{B}} L^2}{4 \sigma (W-w)^2} \frac{1}{(|x| + \delta )^2}
= \frac{V^2 \sigma R_{\mathrm{B}}}{4 \ln^2 (W/w)} \frac{1}{(|x| + \delta )^2} , \label{Twedge} 
\end{eqnarray}
where $\delta$ is given by
\begin{eqnarray}
\delta = \frac{w L}{2(W-w)} .
\end{eqnarray}
This is the distance between the centre of the bow tie and the point where the two converging converging sides of the bow tie would meet if $w=0$, i.e. the separation of $O$ and $O^{\prime}$ in Fig.~\ref{newfig1}.
The resistance~(\ref{rbowtie}) only diverges logarithmically as $w \rightarrow 0$,
and this is a major difference as compared to the nanowire~(\ref{rwire}).
For example, it means that $T_{\mathrm{J},\mathrm{bowtie}}$ for fixed $V$ 
diverges as $w \rightarrow 0$, unlike Eq.~(\ref{Tw}).

As with the nanowire, separation of variables~\cite{hahncomment3} is not easily applicable to the bow tie geometry, Fig.~\ref{fig1}(c), because the heat equation is inhomogeneous~(\ref{Twedge}). To proceed, we derive an effective one-dimensional heat equation assuming homogeneous Joule heating and using separation of variables. We then re-insert the inhomogeneous Joule heating term~(\ref{Twedge}) into the one-dimensional equation, and we compare these solutions with numerical solutions~\cite{fem} of the two-dimensional equation~(\ref{he2}). 

To derive an effective one-dimensional equation, we write the diffusion equation~(\ref{he2}) using polar coordinates $r$ and $\phi$:
\begin{eqnarray}
\frac{\partial^2 \tilde{T}(r,\phi)}{\partial r^2} + \frac{1}{r} \frac{\partial \tilde{T}(r,\phi)}{\partial r} + \frac{1}{r^2}\frac{\partial^2 \tilde{T}(r,\phi)}{\partial \phi^2} + \frac{T_0 + T_{\mathrm{J}} - \tilde{T}(r,\phi)}{L_{\mathrm{H}}^2} = 0 , \label{mbe1}
\end{eqnarray}
where we assume that the Joule heating term, $T_{\mathrm{J}}$, is constant in space.
Notation $\tilde{T}$ is used to specify the solution of this equation in polar coordinates.
The polar coordinates $(r,\phi)$ are measured from the point where the two converging converging sides of the bow tie would meet if $w=0$, e.g. point $O^{\prime}$ in Fig.~\ref{newfig1} is used for the right side of the sample.
We apply separation of variables~\cite{hahncomment3}.
With boundary conditions~(\ref{bcy}) at the edges, $\partial T / \partial \phi = 0$, the heat equation~(\ref{mbe1}) is satisfied by a $\phi$-independent solution which obeys an effective one-dimensional equation,
\begin{eqnarray}
\frac{d^2 \tilde{T}(r)}{d r^2} + \frac{1}{r} \frac{d \tilde{T}(r)}{d r} + \frac{T_0 + T_{\mathrm{J}} - \tilde{T}(r)}{L_{\mathrm{H}}^2} = 0 . \label{mbe2}
\end{eqnarray}
This is the modified Bessel equation of zero order. Note that it has been previously used to model heat transport in annular fins attached to tubes or rods~\cite{acostaiborra09},
and for heating at a localized point source in graphene due either to an applied laser beam or the presence of defects~\cite{cai10,choi17}. In the latter case, $T_{\mathrm{J}} (\mathbf{r})$ had the form of an isotropic Gaussian distribution sharply decaying from the origin in two dimensions, as opposed to the inverse square dependence~(\ref{Twedge}) of Joule heating along a bow tie constriction.

Along the $x$-axis (i.e. $y=0$), where $x$ is the longitudinal coordinate of the bow tie system, the radial coordinate $r = |x| + \delta$ and $r \geq \delta$,
Fig.~\ref{newfig1}.
Overall, we write
\begin{eqnarray}
T (x) &=&  \tilde{T}(- x + \delta) \, \Theta (-x) + \tilde{T}(x + \delta) \, \Theta (x) , \label{xr}
\end{eqnarray}
where $\tilde{T}(r)$ is the solution of Eq.~(\ref{mbe2}) with $r>0$.

For homogeneous $T_{\mathrm{J}}$, the effective heat diffusion equation for the bow tie~(\ref{mbe2}) is the equation for modified Bessel equations of zero order $I_0$ and $K_0$ with general solution~\cite{hahncomment2},
\begin{eqnarray}
\tilde{T}(r) = T_0 + T_{\mathrm{J}} + A I_0 \! \left( \frac{r}{L_{\mathrm{H}}} \right) + B K_0 \! \left( \frac{r}{L_{\mathrm{H}}} \right) , \label{tr1}
\end{eqnarray}
for arbitrary $A$, $B$.
In general, on applying boundary conditions at $x=0$ or $x = \pm L/2$, coefficients $A$ and $B$ will be expressed in terms of the values of the Bessel functions and their derivatives at these points. We use the following notation for the resulting constants:
\begin{eqnarray*}
i_0 &=& I_0 \!\left(\!\frac{\delta}{L_{\mathrm{H}}}\!\right) ; \quad
i_{\infty} = I_0 \!\left(\!\frac{L+2\delta}{2L_{\mathrm{H}}}\!\right) ; \quad i_1 = I_1 \!\left(\!\frac{\delta}{L_{\mathrm{H}}}\!\right) , \\
k_0 &=& K_0 \!\left(\!\frac{\delta}{L_{\mathrm{H}}}\!\right) ; \quad
k_{\infty} = K_0 \!\left(\!\frac{L+2\delta}{2L_{\mathrm{H}}}\!\right) ; \quad
k_1 = K_1 \!\left(\!\frac{\delta}{L_{\mathrm{H}}}\!\right) . 
\end{eqnarray*}

We apply Dirichlet boundary conditions~(\ref{bcx}) at the ends of the device, $x = \pm L/2$, and we use continuity of $T(x)$ and $dT(x)/dx$ at $x=0$. Application of these conditions introduces a small error because they are applied along a straight vertical line at constant $x$ rather than a curved line at constant $r$. This error is negligible for $w / \delta \ll 1$ (i.e. $L \gg W-w$).
The calculation can be simplified by using inversion symmetry in the $x$ direction. This means that the solution must be an even function of $x$, and the boundary conditions can be written as
\begin{eqnarray}
\tilde{T}(r = \tfrac{L}{2} + \delta) = T_0 ; \qquad \left. \frac{d\tilde{T}}{dr} \right|_{r = \delta} = 0 . \label{nbc}
\end{eqnarray}
Applying these to the general solution~(\ref{tr1}) and writing in terms of $x$ using Eq.~(\ref{xr}) gives
\begin{eqnarray}
\!\!\!\!\!\! T (x) &=& T_0 + T_{\mathrm{J}} + A I_0 \! \left( \!\frac{|x| + \delta}{L_{\mathrm{H}}} \!\right) + B K_0 \! \left( \!\frac{|x| + \delta}{L_{\mathrm{H}}} \!\right) \!\! , \\
A &=& - \frac{k_1 T_{\mathrm{J}}}{i_1 k_{\infty} + i_{\infty} k_1} ; \qquad
B = - \frac{i_1 T_{\mathrm{J}}}{i_1 k_{\infty} + i_{\infty} k_1} .
\end{eqnarray}

Now we consider the solution of the effective one-dimensional heat equation for the bow tie constriction~(\ref{mbe2}) with inhomogeneous Joule heating~(\ref{Twedge}) and boundary conditions~(\ref{nbc}). This is done using a Green's functions approach~\cite{greensfunctions}.
For arbitrary Joule heating $T_{\mathrm{J}} (r)$ and arbitrary boundary conditions, we find the general solution to be
\begin{eqnarray}
\tilde{T}(r) = T_0 + \tilde{A} I_0 ( \rho ) + \tilde{B} K_0 ( \rho )  + g_2 (\rho) \int_a^\rho \frac{g_1 (\zeta) T_{\mathrm{J}} (\zeta)}{\varpi (\zeta)} d\zeta + g_1 (\rho) \int_\rho^b \frac{g_2 (\zeta) T_{\mathrm{J}}(\zeta)}{\varpi (\zeta)} d\zeta , \nonumber
\end{eqnarray}
for arbitrary $\tilde{A}$, $\tilde{B}$. The dimensionless longitudinal variable is $\rho = r / L_{\mathrm{H}}$, and it takes values $a = \delta / L_{\mathrm{H}}$ at the centre and $b = (L/2 + \delta)/ L_{\mathrm{H}}$ at the end of the device. The basis functions and Wronskian are given by
\begin{eqnarray}
g_1 (\rho) &=& k_0 I_0 (\rho) - i_0 K_0 (\rho) , \\
g_2 (\rho) &=& k_{\infty} I_0 (\rho) - i_{\infty} K_0 (\rho) , \\
\varpi (\rho) &=& g_1^{\prime} (\rho) g_2 (\rho) - g_1 (\rho) g_2^{\prime} (\rho) .
\end{eqnarray}
With boundary conditions~(\ref{nbc}), the solution is
\begin{eqnarray}
\tilde{T}(r) &=& T_0  - \left[ k_{\infty} I_0 (\rho) - i_{\infty} K_0 (\rho) \right] \frac{(k_0 i_1 + i_0 k_1)}{(k_{\infty} i_1 + i_{\infty} k_1)} \int_a^b \frac{g_2 (\zeta) T_{\mathrm{J}}(\zeta)}{\varpi (\zeta)} d\zeta \nonumber \\
&& \qquad \qquad + g_2 (\rho) \!\int_a^\rho \!\frac{g_1 (\zeta) T_{\mathrm{J}}(\zeta)}{\varpi (\zeta)} d\zeta + g_1 (\rho) \! \int_\rho^b \! \frac{g_2 (\zeta) T_{\mathrm{J}}(\zeta)}{\varpi (\zeta)} d\zeta . \label{twedge}
\end{eqnarray}
and the maximum temperature is given by
\begin{eqnarray}
\tilde{T}(\delta) = T_0 - \frac{(k_{\infty} i_0 - i_{\infty} k_0)(k_0 i_1 + i_0 k_1)}{k_{\infty} i_1 + i_{\infty} k_1} \int_a^b \frac{g_2 (\zeta) T_{\mathrm{J}}(\zeta)}{\varpi (\zeta)} d\zeta . \label{tzbt}
\end{eqnarray}

The temperature profile~(\ref{twedge}) is plotted in Fig.~\ref{fig2}(e),(f) with $T_{\mathrm{J}}$ given by Eq.~(\ref{Twedge}) for different values of $L_{\mathrm{H}}$ (solid lines), and compared with numerical data (symbols) from the finite element method solution~\cite{fem} of the two-dimensional equation~(\ref{he2}).
The solution~(\ref{twedge}) is generally in very good agreement with the numerics.
Dependence of the maximum temperature, $T(0)$, as a function of
$L_{\mathrm{H}}$~\cite{rbcomment}
is shown in Figure~\ref{fig3}(b) where the symbols show numerical data~\cite{fem} and the solid line is Eq.~(\ref{tzbt}). For $\kappa \rightarrow 0$, $T(0) = T_0 + T_{\mathrm{J},\mathrm{bowtie}} (x=0)$ is determined by the Joule heating temperature~(\ref{Twedge}) 
which, for constant $V$, is given by
\begin{eqnarray}
T(0) \rightarrow T_0 + \frac{V^2 \sigma R_{\mathrm{B}} (W-w)^2}{w^2 L^2 \ln^2 (W/w)}
\quad \text{as} \quad \kappa \rightarrow 0 . \label{limit1}
\end{eqnarray}
In the opposite limit, $\kappa \rightarrow \infty$, then $T(0) = T_0$.

Dependence of $T(0)$ as a function of width $w$ is shown in Fig.~\ref{fig4}(c) for constant current $I$ and in Fig.~\ref{fig4}(d) for constant potential difference $V$.
Symbols show numerical data~\cite{fem} and the solid lines are Eq.~(\ref{tzbt}).
As for the nanowire, for $w = W$, the device is simply a rectangle, Eq.~(\ref{rtemp}), with $T (0) = T_0 + T_{\mathrm{J}} - T_{\mathrm{J}}/\cosh (L/(2L_{\mathrm{H}}))$ and $T_{\mathrm{J}} = V^2 \sigma R_{\mathrm{B}} / L^2$ (for constant $V$).

For constant potential difference, Fig.~\ref{fig4}(d), and $w \approx 0$, the integral in the maximum temperature~(\ref{tzbt}) is dominated by the lower limit $a = \delta / L_{\mathrm{H}} \approx 0$ and we approximate
\begin{eqnarray*}
T(0) &\approx& T_0 + T_{\mathrm{J},0} \int_{\delta / L_{\mathrm{H}}}^{\infty} \frac{K_0 (\zeta)}{\zeta} d\zeta , \\
&\approx& T_0 + T_{\mathrm{J},0} \left[ 
\frac{1}{2} \ln^2 \! \left( \frac{2L_{\mathrm{H}}}{\delta} \right) 
- \gamma \ln \left( \frac{L_{\mathrm{H}}}{\delta} \right)
\right] ,
\end{eqnarray*}
where $T_{\mathrm{J},0} = V^2 \sigma R_{\mathrm{B}}/[4 L_{\mathrm{H}}^2 \ln^2 (W/w)]$,
and $\gamma \approx 0.5772157 \ldots$ is the Euler-Mascheroni  constant.
This describes the logarithmic behaviour for $w \approx 0$, Fig.~\ref{fig4}(d). The temperature actually goes to a finite value as $w \rightarrow 0$ (for non-zero $L_{\mathrm{H}}$) because logarithmically divergent factors in $T_{\mathrm{J},0}$ and in the integral cancel:
\begin{eqnarray}
T(0) \rightarrow T_0 + \frac{V^2 \sigma R_{\mathrm{B}}}{8 L_{\mathrm{H}}^2} \quad
\text{as} \quad w \rightarrow 0 . \label{limit2}
\end{eqnarray}

For constant current, Fig.~\ref{fig4}(c), the temperature diverges as $w \rightarrow 0$ because the current density diverges. However, the divergence is only logarithmic in comparison to $w^{-2}$ for the nanowire, and this may be shown analytically with a calculation similar to Eq.~(\ref{limit2}). Results for the maximum temperature $T(0)$ in various limits are summarised in table~\ref{tablevi}. 
As with the nanowire, the dependence on $R_{\mathrm{B}}$ is distinct from that of $\kappa$: when $R_{\mathrm{B}} \rightarrow 0$, heat can immediately dissipate into the environment and $T(0) \rightarrow T_0$, whereas $R_{\mathrm{B}} \rightarrow \infty$ increases the temperature $T(0)$ at the centre of the device.
For the bow tie, table~\ref{tablevi} shows that the maximum temperature has the same parameter dependence for $R_{\mathrm{B}} \rightarrow \infty$ as for $\omega \rightarrow 0$. This is because both behaviours are driven by the logarithmically divergent temperature profile at the narrowest part of the bow tie.

\section{Temperature dependent thermal conductivity}

We have assumed electrical and thermal conductivities to be independent of temperature, which is an approximation. For example, graphene has a weak, but non-zero temperature dependence of its thermal conductivity at room temperature~\cite{sachat21,dorgan13,pop12}. To account for this, one could incorporate the temperature dependences of scattering rates into the conductivities~\cite{nika12,dorgan13,antoulinakis16}.
As an example, we use the phenomenological form of the temperature dependence of the thermal conductivity proposed in Ref.~\cite{dorgan13},
\begin{eqnarray}
\kappa (T) = \kappa_0 \bigg( \frac{T_0}{T} \bigg)^{\gamma} , \label{kappat}
\end{eqnarray}
where $\kappa_0$ is the thermal conductivity at room temperature $T_0$, and $\gamma$ is a fitting parameter. Ref.~\cite{dorgan13} finds $\gamma \approx 1.9$ and $\gamma \approx 1.7$ by fitting to experimental data for graphene devices produced by exfoliation and chemical vapor deposition, respectively.

Eq.~(\ref{kappat}) describes a decrease of thermal conductivity with increasing temperature. This should lead to a reduction in the effective healing length in the hottest part of the sample (i.e. the centre), producing a further increase in temperature there because $T({\bf r}) \rightarrow T_0 + T_{\mathrm{J}}({\bf r})$ as
$\kappa \rightarrow 0$ (Eq.~(\ref{lhlimit})). To verify this expectation, we consider the steady state heat equation~(\ref{he1}) including $\kappa (T)$, and write this as
\begin{eqnarray}
\nabla \cdot \left[ \frac{\nabla t ({\bf r})}{(1 + \chi t({\bf r}))^{\gamma}} \right]
- \frac{t({\bf r})}{L_{\mathrm{H}}^2}
+ \frac{T_{\mathrm{J}} ({\bf r})}{T_{\mathrm{J},\mathrm{rect}} L_{\mathrm{H}}^2} = 0 , \label{nlhe}
\end{eqnarray}
where
\begin{eqnarray}
t({\bf r}) = \frac{T({\bf r}) - T_0}{T_{\mathrm{J},\mathrm{rect}}} ; \qquad
\chi = \frac{T_{\mathrm{J},\mathrm{rect}}}{T_0} ,
\end{eqnarray}
and $L_{\mathrm{H}} = \sqrt{\kappa_0 R_B}$. As previously, we consider the reduced temperature $t({\bf r})$ normalised by $T_{\mathrm{J},\mathrm{rect}}$, Eq.~(\ref{trect}). As this is a nonlinear equation, the potential difference (or current) appears via the additional parameter $\chi$ which describes the level of Joule heating as compared to room temperature.

\begin{figure}[t]
\includegraphics[scale=0.5]{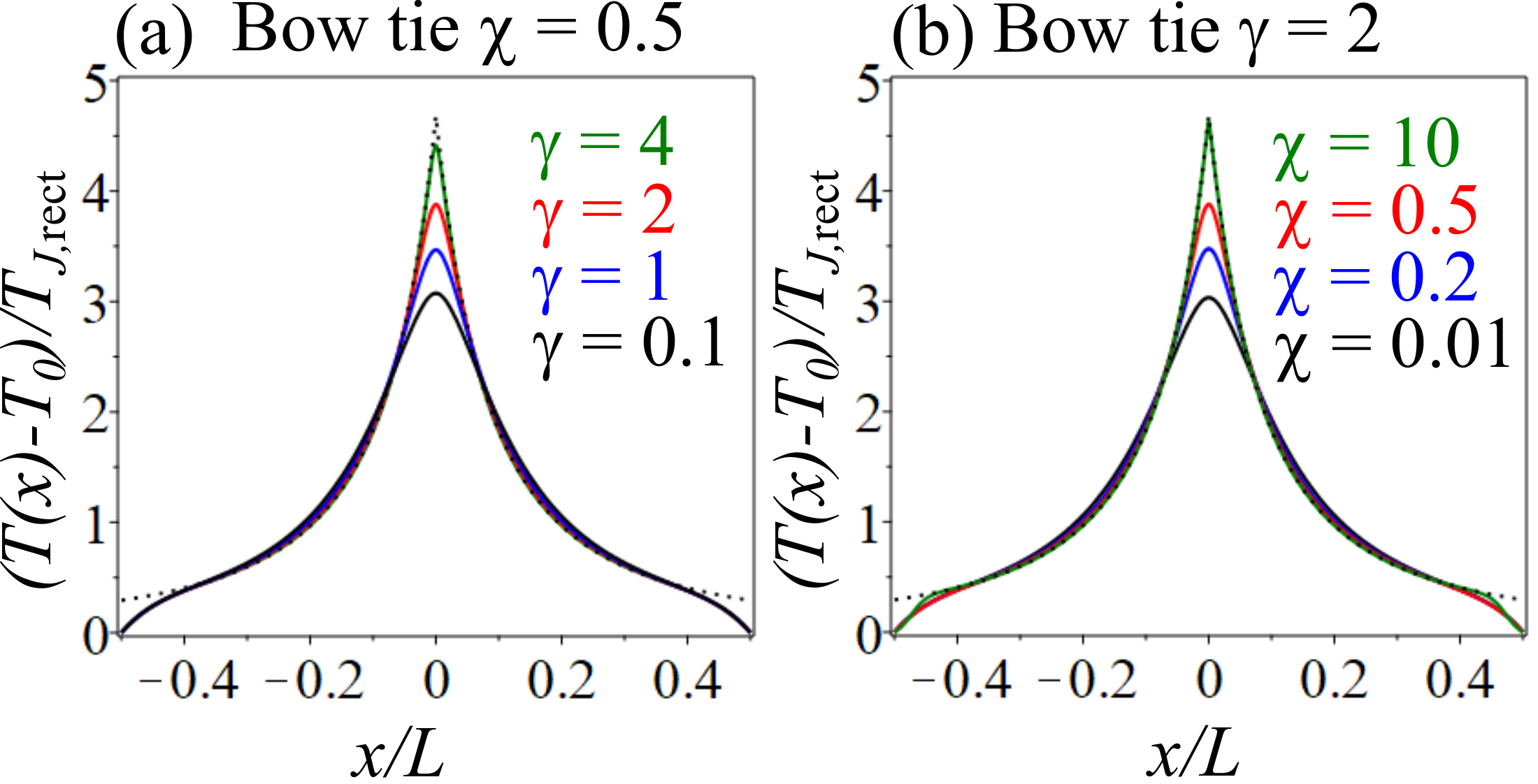}
\caption{The temperature profile $T(x) - T_0$ in the longitudinal direction for a bow tie device taking into account temperature dependent thermal conductivity~(\ref{kappat}).
The maximum width is $W = 0.5L$, the minimum width is $w = 0.125L$, and the thermal healing length is $L_{\mathrm{H}} = 0.05L$.
The plots are normalised by the Joule temperature for a rectangle $T_{\mathrm{J},\mathrm{rect}} = V^2 \sigma R_{\mathrm{B}} / L^2$.
Solid lines show numerical data from the finite element method solution~\cite{fem} of the two-dimensional nonlinear equation~(\ref{nlhe}).
(a) is for fixed parameter $\chi = 0.5$ with $\gamma = 0.1, 1, 2, 4$ (from bottom to top),
(b) is for fixed parameter $\gamma = 2$ with $\chi = 0.01, 0.2, 0.5, 10$ (from bottom to top).
Dotted lines show the Joule heating profile~(\ref{lhlimit}) that is valid in the limit
$L_{\mathrm{H}} \rightarrow 0$~\cite{rbcomment}.
}\label{fig5}
\end{figure}

We numerically solve the two-dimensional nonlinear heat equation~(\ref{nlhe}) with boundary conditions~(\ref{bcx},\ref{bcy}) using the finite element method~\cite{fem}. As an example, the temperature profile for a bow tie device is shown in Fig.~\ref{fig5} for the same parameters as in Fig.~\ref{fig2}(f), namely, maximum width $W = 0.5L$, minimum width $w = 0.125L$, and $L_{\mathrm{H}} = 0.05L$.
Fig.~\ref{fig5}(a) shows the effect of changing parameter $\gamma$ for fixed $\chi$, 
Fig.~\ref{fig5}(b) is changing $\chi$ for fixed $\gamma$.
In both cases, increasing nonlinearity (i.e. larger $\gamma$ or larger $\chi$) leads to an interpolation of the temperature near the centre between the linear case (solid line in Fig.~\ref{fig2}(f)) 
and the limit of $T({\bf r}) = T_0 + T_{\mathrm{J}}({\bf r})$ (dotted line in Fig.~\ref{fig2}(f)), in agreement with our initial expectation.
Although our analytic formulae in previous sections assume temperature independent conductivities, this example demonstrates that they may be used as the basis for qualitative understanding of more complicated situations including temperature dependent parameters $\kappa$, $\sigma$ and $R_{\mathrm{B}}$.

\section{Conclusions}

We modelled heat transport and the resulting spatial temperature profile in the presence of Joule self-heating due to electrical current in nanoconstrictions in monolayer 2D materials. In particular, we considered two contrasting geometries, Fig.~\ref{fig1}: the nanowire has a constant width $w$ whereas the bow tie has a variable width (with a minimum of $w$).
Different shapes produce different parameter dependences of the Joule self-heating and the resultant temperature, and analytic results for the maximum device temperature in various limits are summarised in table~\ref{tablevi}. 
When the system is a poor thermal conductor ($\kappa \rightarrow 0$), the maximum temperature is determined by the Joule self-heating at the narrowest point of width $w$.
For fixed potential difference $V$, 
this is independent of $w$ for the nanowire (for small $w$), but diverges as $w^{-2}$ for the bow tie.
For fixed current $I$, however, the temperature diverges as $w^{-2}$ in both devices.
More generally, for a good thermal conductor, the maximum temperature is finite for fixed $V$ as $w \rightarrow 0$, but diverges for fixed $I$: as $w^{-2}$ for the nanowire, but only logarithmically for the bow tie.

Electrical and thermal transport were considered in the classical, diffusive regime, applicable to experiments at room temperature or above, and with spatial dimensions typically of the order of~$100\,$nm or above~\cite{liao11,dorgan13,zolotavin17,harzheim18,evangeli21}. At lower temperatures, with smaller devices, and depending on sample quality, it is necessary to consider ballistic transport~\cite{chen00,bae13,kaiser17,li19,sachat21} and, possibly, quantum interference effects. 
Joule heating was considered in the linear electrical transport regime, but the parameter dependences of the Joule heating temperature $T_{\mathrm{J}}$ could be modified to take into account non-linear $I(V)$ characteristics.
It is also possible to adapt the effective one-dimensional equations to describe local heating, e.g. due to a scanning thermal microscope~\cite{harzheim18}, by introducing either discontinuous matching conditions for $dT/dx$ or a local Joule heating term~\cite{cai10,choi17}.

\section*{Data Availability Statement}

The data that support the findings of this study are available in Lancaster University Research Directory at https://doi.org/10.17635/lancaster/researchdata/470.


\begin{thebibliography}{99}

\bibitem{pop10}
E. Pop,
Nano.\ Res.\ {\bf 3}, 147 (2010).

\bibitem{fu20}
Y. Fu, J. Hansson, Y. Liu, S. Chen, A. Zehri, M. K. Samani, N. Wang, Y. Ni, Y. Zhang, Z.-B. Zhang, Q. Wang, M. Li, H. Lu, M. Sledzinska, C. M. Sotomayor Torres, S. Volz, A. A. Balandin, X. Xu, and J. Liu,
2D Mater.\ {\bf 7}, 012001 (2020)

\bibitem{lewis21}
J. S. Lewis, T. Perrier, Z. Barani, F. Kargar and A. A Balandin,
Nanotechnology {\bf 32}, 142003 (2021).

\bibitem{sachat21}
A. El Sachat, F. Alzina, C. M. Sotomayor Torres, and E. Chavez-Angel,
Nanomaterials {\bf 11}, 175 (2021).

\bibitem{ferrari15}
A. C. Ferrari et al,
Nanoscale {\bf 7}, 4598 (2015).

\bibitem{debroy20}
S. Debroy, S. Sivasubramani, G. Vaidya, S. G. Acharyya, and A. Acharyya,
Sci.\ Rep.\ {\bf 10}, 6240 (2020).

\bibitem{son21}
M. Son, J. Jang, Y. Lee, J. Nam, J. Y. Hwang, I. S. Kim, B. H. Lee, M.-H. Ham, and S.-S. Chee,
NPJ 2D Mater.\ Appl.\ {\bf 5}, 41 (2021).

\bibitem{zong20}
P.-a. Zong, J. Liang, P. Zhang, C. Wan, Y. Wang, and K. Koumoto,
ACS Appl.\ Energy Mater.\ {\bf 3} 2224 (2020).

\bibitem{wang20}
J. Wang , Z. Xie and J. T. W. Yeow,
Mater.\ Res.\ Express {\bf 7}, 112001 (2020).

\bibitem{pallecchi20}
I. Pallecchi, N. Manca, B Patil, L. Pellegrino, and D. Marr\'e,
Nano Futures {\bf 4} 032008 (2020).

\bibitem{zhao21}
J. Zhao, D. Ma, C. Wang, Z. Guo, B. Zhang, J. Li, G. Nie, N. Xie, and H. Zhang,
Nano Research {\bf 14}, 897 (2021).

\bibitem{jeong14}
W. Jeong, K. Kim, Y. Kim, W. Lee, and P. Reddy,
Sci.\ Rep.\ {\bf 4}, 4975 (2014).

\bibitem{island14}
J. O. Island, A. Holovchenko, M. Koole, P. F. A. Alkemade, M. Menelaou, N. Aliaga-Alcalde, E. Burzur\'i, and H. S. J. van der Zant,
J. Phys.: Condens.\ Matter {\bf 26}, 474205 (2014).

\bibitem{abbassi17}
M. El Abbassi, L. P\'osa, P. Makk, C. Nef, K. Thodkar, A. Halbritter, and M. Calame,
Nanoscale {\bf 9}, 17312 (2017).

\bibitem{gu18}
C. Gu, D. Su, C. Jia, S. Rena, and X. Guo,
RSC Adv.\ {\bf 8}, 6814 (2018).

\bibitem{celis16}
A. Celis, M. N. Nair, A. Taleb-Ibrahimi, E. H. Conrad, C. Berger, W. A. de Heer and A. Tejeda,
J.\ Phys.\ D: Appl.\ Phys.\ {\bf 49}, 143001 (2016).

\bibitem{saraswat21}
V. Saraswat, R. M. Jacobberger, and M. S. Arnold,
ACS Nano {\bf 15}, 3674 (2021).

\bibitem{liao11}
A. Liao, J. Z. Wu, X. Wang, K. Tahy, D. Jena, H. Dai, and E. Pop,
Phys.\ Rev.\ Lett.\ {\bf 106}, 256801 (2011).

\bibitem{zolotavin17}
P. Zolotavin, C. I. Evans, and D. Natelson,
Nanoscale {\bf 9}, 9160 (2017).

\bibitem{harzheim18}
A. Harzheim, J. Spiece, C. Evangeli, E. McCann, V. Falko, Y. Sheng, J. H. Warner, G. A. D. Briggs, J. A. Mol, P. Gehring, and O. V. Kolosov,
Nano Lett.\ {\bf 18}, 7719 (2018).

\bibitem{evangeli21}
C. Evangeli, E. McCann, J. L. Swett, S. Tewari, X. Bian, J. O. Thomas, G. A. D. Briggs, O. V. Kolosov, and J. A. Mol,
Carbon {\bf 178}, 632 (2021).

\bibitem{shi09}
L. Shi, J. Zhou, P. Kim, A. Bachtold, A. Majumdar, and P. L. McEuen,
J.\ Appl.\ Phys.\ {\bf 105}, 104306 (2009).

\bibitem{puyoo11}
E. Puyoo, S. Grauby, J.-M. Rampnoux, E. Rouvi\`ere, and S. Dilhaire,
J.\ Appl.\ Phys.\ {\bf 109}, 024302 (2011).

\bibitem{menges16}
F. Menges, P. Mensch, H. Schmid, H. Riel, A. Stemmer, and B. Gotsmann,
Nat.\ Commun.\ {\bf 7}, 10874 (2016).

\bibitem{sachet17}
A. El Sachat, J. S. Reparaz, J.Spiece, M. I. Alonso, A. R. Go\~ni, M. Garriga, P. O. Vaccaro, M. R. Wagner, O. V. Kolosov, and C. M. Sotomayor Torres,
Nanotechnology {\bf 28} 505704 (2017).

\bibitem{gachter20}
N. G\"achter, F. K\"onemann, M. Sistani, M. G. Bartmann, M. Sousa, P. Staudinger, A. Lugstein, and B. Gotsmann,
Nanoscale {\bf 12}, 20590 (2020).

\bibitem{pumarol12}
M. E. Pumarol, M. C. Rosamond, P. Tovee, M. C. Petty, D. A. Zeze, V. Falko, and O. V. Kolosov,
Nano Lett. {\bf 12}, 2906 (2012).

\bibitem{menges13}
F. Menges, H. Riel, A. Stemmer, C. Dimitrakopoulos, and B. Gotsmann,
Phys.\ Rev.\ Lett.\ {\bf 111}, 205901 (2013).

\bibitem{yoon14}
K. Yoon, G. Hwang, J. Chung, H. G. Kim, O. Kwon, K. D. Kihm, J. S. Lee,
Carbon {\bf 76}, 77 (2014).

\bibitem{tortello16}
M. Tortello, S. Colonn, M. Bernal, J. Gomez, M. Pavese, C. Novara, F. Giorgis, M. Maggio, G. Guerra, G. Saracco, R. S. Gonnelli, and A. Fina,
Carbon {\bf 109}, 390 (2016).

\bibitem{choi17}
D. Choi, N. Poudel, S. B. Cronin, and L. Shi,
Appl.\ Phys.\ Lett.\ {\bf 110}, 073104 (2017).

\bibitem{yalon17}
E. Yalon, C. J. McClellan, K. K. H. Smithe, M. M. Rojo, R. L. Xu, S. V. Suryavanshi, A. J. Gabourie, C. M. Neumann, F. Xiong, A. B. Farimani, and E. Pop,
Nano Lett.\ {\bf 17}, 3429 (2017).

\bibitem{tortello19}
M. Tortello, I. Pasternak, K. Zeranska-Chudek, W. Strupinski, R. S. Gonnelli, and A. Fina,
ACS Appl.\ Nano Mater.\ {\bf 2}, 2621 (2019).

\bibitem{sachat19}
A. El Sachat, F. K\"onemann, F. Menges, E. Del Corro, J. A. Garrido, C. M. Sotomayor Torres, F. Alzina,
and B Gotsmann,
2D Mater.\ {\bf 6}, 025034 (2019)

\bibitem{yasaei19}
P. Yasaei, A. A. Murthy, Y. Xu  Roberto dos Reis, G. S. Shekhawat, and V. P. Dravid,
Adv.\ Mater.\ {\bf 31}, e1808244 (2019).

\bibitem{evangeli19}
C. Evangeli, J. Spiece, S. Sangtarash, A. J. Molina‐Mendoza, M. Mucientes, T. Mueller, C. Lambert, H. Sadeghi, and O. Kolosov,
Adv.\ Electron.\ Mater.\ {\bf 5} 1900331 (2019).

\bibitem{harzheim20}
A. Harzheim, C. Evangeli, O. V Kolosov, and P. Gehring,
2D Mater.\ {\bf 7}, 041004 (2020).

\bibitem{dorgan13}
V. E. Dorgan, A. Behnam, H. J. Conley, K. I. Bolotin, and E. Pop,
Nano Lett.\  {\bf 13}, 4581 (2013).

\bibitem{durkan99}
C. Durkan, M. A. Schneider, and M. E. Welland,
J.\ Appl.\ Phys.\ {\bf 86}, 1280 (1999).

\bibitem{kuroda05}
M. A. Kuroda, A. Cangellaris, and J.-P. Leburton,
Phys.\ Rev.\ Lett.\ {\bf 95}, 266803 (2005).

\bibitem{pop05}
E. Pop, D. Mann, J. Reifenberg, K. E. Goodson, and H. J. Dai,
in IEEE International Electron Devices Meeting (IEDM), Washington, DC
(IEEE, New York, 2005, pp. 253–256).

\bibitem{pop07}
E. Pop, D. A. Mann, K. E. Goodson, and H. Dai,
J.\ Appl.\ Phys.\ {\bf 101}, 093710 (2007).

\bibitem{hunley13}
D. P. Hunley, S. L. Johnson, R. L. Flores, A. Sundararajan, and D. R. Strachan,
J.\ Appl.\ Phys.\ {\bf 113}, 234306 (2013).

\bibitem{chandran15}
K. S. R. Chandran,
Int.\ J.\ Heat Mass Transf.\ {\bf 88}, 14 (2015).

\bibitem{pu19}
P. Pu, M. Tang, and J. Mao,
2019 Cross Strait Quad-Regional Radio Science and Wireless Technology Conference (CSQRWC), pp. 1-2 (2019).

\bibitem{fangohr11}
H. Fangohr, D. S. Chernyshenko, M. Franchin, T. Fischbacher, and G. Meier,
Phys.\ Rev.\ B {\bf 84}, 054437 (2011).

\bibitem{islam13}
S. Islam, Z. Li, V. E. Dorgan, M.-H. Bae, and E. Pop,
IEEE Electron Device Lett.\ {\bf 34},  (2013)

\bibitem{ramos15}
E. Ramos, C. L\'opez, J. Akerman, M. Mu\~noz, and J. L. Prieto,
Phys.\ Rev.\ B {\bf 91}, 214404 (2015).

\bibitem{hadeed07}
F. O. Hadeed and C. Durkan,
Appl.\ Phys.\ Lett.\ {\bf 91}, 123120 (2007).

\bibitem{xiang14}
A. Xiang, S. Hou, and J. Liao,
Appl.\ Phys.\ Lett.\ {\bf 104}, 223113 (2014).

\bibitem{sawtelle19}
S. D. Sawtelle and M. A. Reed,
Phys.\ Rev.\ B {\bf 99}, 054304 (2019).

\bibitem{ramachandran05}
G. K. Ramachandran, M. D. Edelstein, D. L. Blackburn, J. S. Suehle, E. M. Vogel, and C. A. Richter,
Nanotechnology {\bf 16}, 1294 (2005).

\bibitem{lu10}
Y. Lu, B. Goldsmith, D. R. Strachan, J. H. Lim, Z. Luo, and A. T. C. Johnson,
Small {\bf 6}, 2748 (2010).

\bibitem{fem} We numerically solve the two-dimensional heat equation~(\ref{he2}) using boundary conditions~(\ref{bcx},\ref{bcy}). This is done using finite element method calculations in the MathWorks Partial Differential Equation Toolbox \url{https://www.mathworks.com/help/pde/index.html}.

\bibitem{chiang02}
T.-Y. Chiang, K. Banerjee, and K. C. Saraswat,
IEEE Electron Device Lett.\ {\bf 23}, 31 (2002).

\bibitem{hahn13}
{\em Heat Conduction}
by D. W. Hahn and M. N. \"{O}zisik,
(John Wiley \& Sons, Inc., Hoboken, New Jersey, 2012, 3rd Ed.)

\bibitem{hahncomment1} See Eq.~(1-25) in Ref.~\cite{hahn13}. We denote the rate of energy generation as $p$ instead of $g$ to avoid confusion with electrical conductance.

\bibitem{dimensionscomment}
For simplicity, we assume a strictly two-dimensional material. It is straightforward to generalize our approach to take into account a finite thickness by adjusting the physical dimensions of quantities such as (volumetric) rate of energy generation $p$, and electric and heat current densities.

\bibitem{dorgan10}
V. E. Dorgan, M.-H. Bae, and E. Pop,
Appl.\ Phys.\ Lett.\ {\bf 97}, 082112 (2010).

\bibitem{balandin11}
A. A. Balandin,
Nat.\ Mater.\ {\bf 10}, 569 (2011).

\bibitem{pop12}
E. Pop, V. Varshney, A. K. Roy,
MRS Bulletin, vol. 37, pp. 1273-1281 (2012).

\bibitem{mleczko16}
M. J. Mleczko, R. L. Xu, K. Okabe, H.-H. Kuo, I. R. Fisher, H.-S. P. Wong, Y. Nishi, and E. Pop,
ACS Nano {\bf 10}, 7507 (2016).

\bibitem{barcohen15}
A. Bar-Cohen, K. Matin, and S. Narumanchi,
J.\ Electron.\ Packag.\ {\bf 137}, 040803 (2015).

\bibitem{denisov20}
A. O. Denisov, E. S. Tikhonov, S. U. Piatrusha, I. N. Khrapach, F. Rossella, M. Rocci, L. Sorba, S. Roddaro, and V. S Khrapai,
Nanotechnology {\bf 31}, 324004 (2020)

\bibitem{hahncomment3} Separation of variables is described in Chapters 3,4 and 5 of Ref.~\cite{hahn13}.

\bibitem{liao10}
A. Liao, R. Alizadegan, Z.-Y. Ong, S. Dutta, F. Xiong, K. J. Hsia, and E. Pop,
Phys.\ Rev.\ B {\bf 82}, 205406 (2010).

\bibitem{rbcomment}
In Figs.~\ref{fig2},\ref{fig3},\ref{fig4},\ref{fig5}, the vertical scale is normalised by $T_{\mathrm{J},\mathrm{rect}}$, Eq.~(\ref{trect}), which is proportional to the thermal boundary resistance $R_{\mathrm{B}}$. This should be noted when varying the thermal healing length $L_{\mathrm{H}} = \sqrt{\kappa R_{\mathrm{B}}}$.

\bibitem{acostaiborra09}
A. Acosta-Iborra and A. Campo,
Int.\ J.\ Therm.\ Sci.\ {\bf 48}, 773 (2009).

\bibitem{cai10}
W. Cai, A. L. Moore, Y. Zhu, X. Li, S. Chen, L. Shi, and R. S. Ruoff,
Nano Lett.\ {\bf 10}, 1645 (2010)

\bibitem{hahncomment2}
See Eqs.~(2-59) and~(2-60) in Ref.~\cite{hahn13}.

\bibitem{greensfunctions}
{\em Green's functions: construction and applications}
by Y. A. Melnikov and M. Y. Melnikov,
(de Gruyter, Berlin/Boston, 2012).

\bibitem{nika12}
D. L. Nika and A. A Balandin,
J.\ Phys.: Condens.\ Matter {\bf 24}, 233203 (2012). 

\bibitem{antoulinakis16}
F. Antoulinakis, D. Chernin, P. Zhang, and Y. Y. Lau,
J.\ Appl.\ Phys.\ {\bf 120}, 135105 (2016).

\bibitem{chen00}
G. Chen,
J.\ Nanoparticle Res.\ {\bf 2}, 199 (2000).

\bibitem{bae13}
M.-H. Bae, Z. Li, Z. Aksamija, P. N. Martin, F. Xiong, Z.-Y. Ong, I. Knezevic, and E. Pop,
Nat.\ Commun.\ {\bf 4}, 1734 (2013). 

\bibitem{kaiser17}
J. Kaiser, T. Feng, J. Maassen, X. Wang, X. Ruan, and M. Lundstrom,
J.\ Appl.\ Phys.\ {\bf 121}, 044302 (2017).

\bibitem{li19}
Q.-Y. Li, T. Feng, W. Okita, Y. Komori, H. Suzuki, T. Kato, T. Kaneko, T. Ikuta, X. Ruan, and K. Takahashi,
ACS Nano {\bf 13}, 9182 (2019).

\end{thebibliography}
\end{document}